\documentclass[fleqn,10pt]{wlscirep}

\usepackage[upright]{fourier}
\usepackage[utf8]{inputenc}
\usepackage[T1]{fontenc}
\renewcommand{\thesubsection}{\arabic{subsection}}
\usepackage{lineno}

\title{Experimental Demonstration of a Rowland Spectrometer for Spin Waves}

\author[1,2,+]{\'{A}d\'{a}m Papp}
\author[1,+]{Martina Kiechle}
\author[1]{Simon Mendisch}
\author[1]{Valentin Ahrens}
\author[1]{Levent Sahin}
\author[1]{Lukas Seitner}
\author[3]{Wolfgang Porod}
\author[1,2]{Gyorgy Csaba}
\author[1,*]{Markus Becherer}
\affil[1]{Technical University of Munich (TUM), Department of Electrical and Computer Engineering, Munich, Germany}
\affil[2]{P\'{a}zm\'{a}ny P\'{e}ter Catholic University, Faculty of Information Technology and Bionics, Budapest, Hungary}
\affil[3]{University of Notre Dame, Department of Electrical Engineering, Notre Dame, IN, 46556}
\affil[+]{These authors contributed equally to this work.}

\affil[*]{markus.becherer@tum.de}

\begin{abstract}
We experimentally demonstrate the operation of a spin-wave Rowland spectrometer. In the proposed device geometry, spin waves are coherently excited on a diffraction grating and form an interference pattern that spatially separates spectral components of the incoming signal. The diffraction grating was created by focused-ion-beam irradiation, which was found to locally eliminate the ferrimagnetic properties of YIG, without removing the material. We found that in our experiments spin waves were created by an indirect mechanism, by exploiting nonlinear resonance between the grating and the coplanar waveguide. Our work paves the way for complex spin-wave optic devices -- chips that replicate the functionality of integrated optical devices on a chip-scale.
\end{abstract}
\begin{document}

\flushbottom
\maketitle
\thispagestyle{empty}

\section*{Introduction}
\label{Intro}
Information processing in today's computers is done almost exclusively by charges (electric currents). Photons, albeit they are ideal for information transmission, never became a mainstream technology for computing. Despite their numerous advantages, photonic devices have practical limitations: they are challenging to integrate on-chip and optical wavelengths (about a micrometer) are huge compared to nanoscale devices, limiting the scalability of any photonic interference-based device.

Spin waves (aka magnons) are wave-like excitations in ferromagnetic (and ferrimagnetic) materials that travel via coupling between precessing magnetic moments. Their wavelength can be adjusted in a wide range (from several micrometers down to potentially nanometer scale), and they have an electronics-friendly frequency range (1-100 GHz) \cite{ref:persp}. This makes spin waves attractive for on-chip applications, especially in wave-based microwave signal processing. They also interact with each other (scatter), and the resulting nonlinearity may enable general-purpose computation. Spin waves exist only in magnetic media and one needs high-quality materials to achieve ideal conditions for propagation and also carefully designed waveguide structures to launch (and pick up) the waves. It has only recently become possible to demonstrate short-wavelength, long-distance propagation \cite{ref:mingzhong, ref:grundler}, spin-wave equivalents of optical laws \cite{ref:snell} or larger-scale refractive devices using local heating \cite{ref:freymann}. Spin-wave variants of complex optical devices are within reach of experimental demonstrations \cite{ref:albisetti}.

A key element of such a spin-wave optics device is a source which launches coherent spin waves. In the simplest case, the source of spin waves can be a simple coplanar waveguide that is placed atop the magnetic film. For many device constructions, such a simple construction is insufficient. Microwave waveguides alone are fairly inefficient at short spin-wave wavelengths \cite{ref:giovanni} and they are limited to generation of plane wavefronts or curved wavefronts with relatively small curvature. For generation of short-wavelength spin waves or non-planar wavefronts, lithographically patterning the edge of the magnetic film is desirable. A periodically patterned edge can serve both as a wave source and a diffracting element.

The focus of the present paper is the experimental study of such a patterned edge as a spin-wave launcher. We use Focused Ion Beam (FIB) irradiation to write a high-resolution concave grating pattern in an yttrium iron garnet (YIG) thin film. The grating is used in the Rowland spectrometer arrangement, which is frequently used in optical and X-ray spectroscopy \cite{ref:james}. This arrangement does not require a separate lens component, and thus it is ideal for waves with limited propagation length. We experimentally demonstrate that the edge of a FIB-irradiated pattern in YIG generates a coherent spin-wave wavefront. Using time-resolved Magneto Optical Kerr Effect (trMOKE) imaging we found that the diffraction patterns closely match those expected from theory and micromagnetic simulations. An unexpected discovery was that in our experiments spin waves were not primarily excited directly by the field of the waveguide. Instead, spin waves were generated indirectly by the dipole fields of high-amplitude, nonlinear, and long-wavelength standing waves that developed behind the grating in the unirradiated area. This process has a significantly higher efficiency of spin-wave generation at a distance from the waveguide. However, this quasi-homogeneous oscillation behind the grating only forms at a sufficiently high amplitude and only if a single frequency component is applied. 

\section*{Spectral Decomposition with a Concave Diffraction Grating}
\label{sec:SAoperation}
\begin{figure}[ht]
    \includegraphics[width=0.99\textwidth]{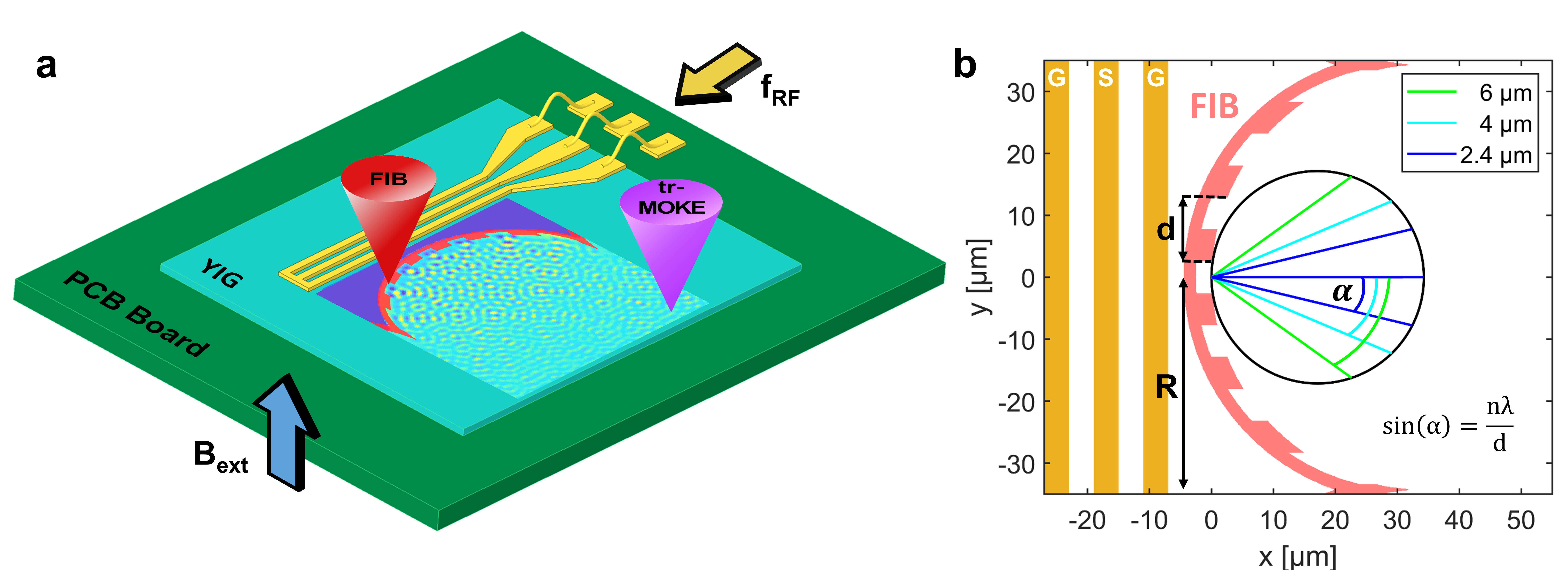}
    \caption{ \textbf{a}) Sketch of the experimental setup, indicating the coplanar waveguide, the FIB irradiated grating, the trMOKE probing laser, and the spin-wave interference pattern. \textbf{b}) Geometry of the curved diffraction grating in the Rowland arrangement. The angle under which the first-order diffraction peak is seen from center of the grating is denoted by $\alpha$.}
    \label{fig:sketch}
\end{figure}

The main component of the proposed device is a concave diffraction grating that acts both as a wave source and as a diffractor. The fabrication of such a grating requires sub-wavelength patterning resolution as the pitch of the grating has to be comparable to the spin-wave wavelength. We use the so-called Rowland arrangement, a detailed description is given in \cite{ref:scirep}. The device generates a spectral decomposition of a time-domain signal by converting temporal frequency components to spatially separated spin-wave intensity peaks. The layout of the fabricated device is shown in Fig. \ref{fig:sketch}. The microwave signal is converted to spin waves by a waveguide antenna. Each frequency component of the signal generates spin waves with corresponding wavelengths. Along the edge of the grating (FIB-irradiated region) the time-varying magnetic field is almost homogeneous, but due to the abrupt parameter change in YIG, every point along the edge acts as a wave source (also described in \cite{ref:kruglyak}). The curved grating not only diffracts different wavelengths to different directions, but also focuses the wavefronts. The drawing of Fig. \ref{fig:sketch}\textbf{b} shows the geometry to determine the diffraction pattern. With a concave grating of radius $R$ and ridge pitch $d$, the diffraction peaks (i.e. wavelength-dependent focal points) will form on a circle with radius $R/2$ drawn tangentially to the grating (Rowland circle). The $n^{th}$-order diffraction angle can be calculated as $\alpha = \arcsin{\left(\frac{n\lambda}{d}\right)}$, where $\lambda$ is the wavelength of the spin wave.

In traditional optical or X-ray Rowland spectrometers the wave source is placed opposing the curved grating, which reflects the waves, acting as a secondary source. Such arrangement would be rather impractical for spin waves: the relatively long path between the source and the diffraction grating will cause much higher attenuation of spin waves. Thus, it is desirable that the grating and the source are in the same structure, i.e. the coherent spin-wave source itself is shaped as a curved diffraction grating. In this geometry, diffraction is caused by the phase difference between waves that originate from the bottom and the top of the ridges. This phase difference will also depend on the ratio of the ridge depth and the wavelength, which does not influence the diffraction angle, but it changes the relative amplitude between diffraction orders. In the designed structure the ridge depth also introduces an amplitude difference between waves that are generated on the top and the bottom of the ridge. The device is thus a combination of an amplitude grating and a phase grating. 

\begin{figure}[ht]    
\centering
    \includegraphics[width=0.99\textwidth]{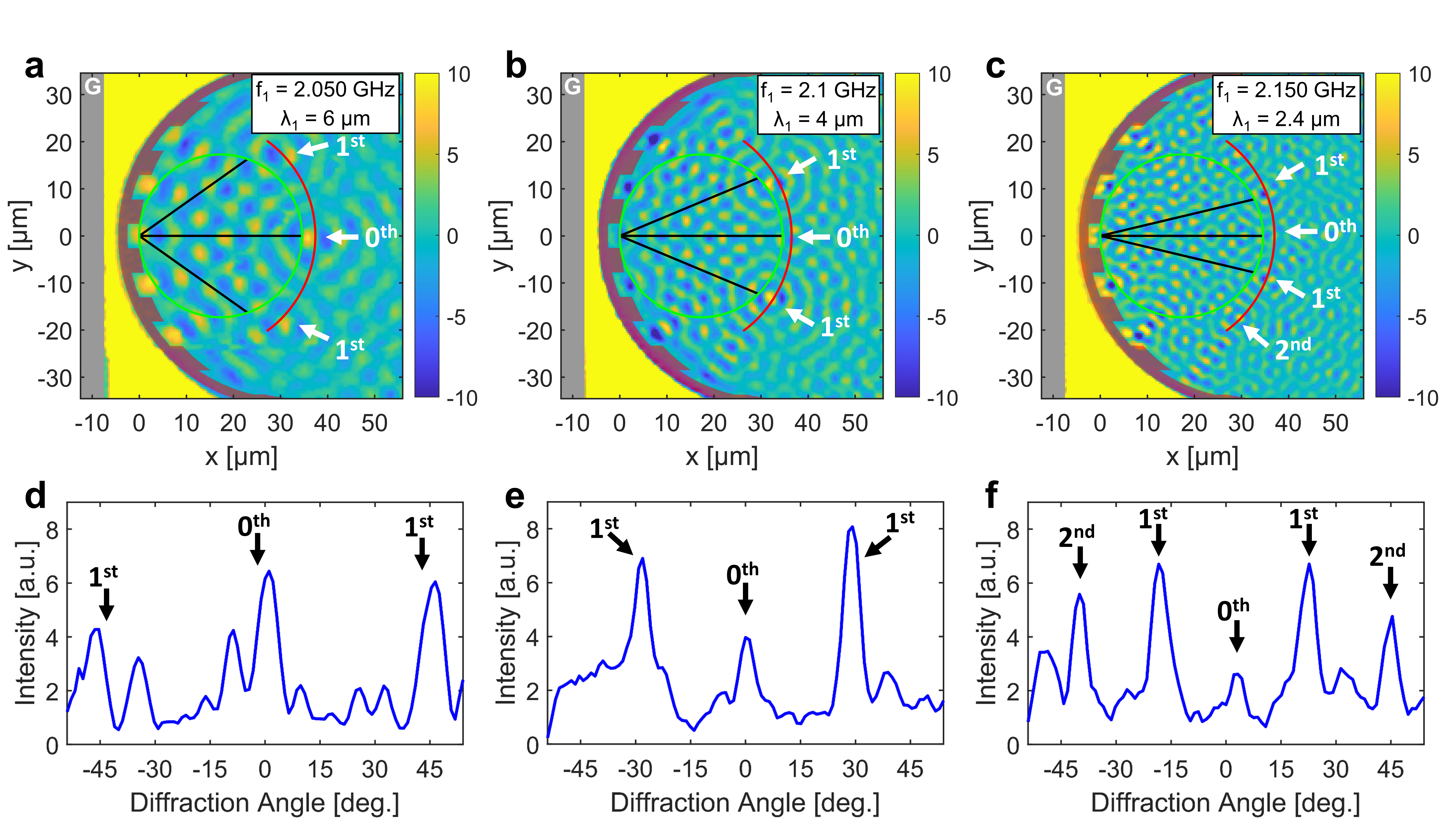}
\caption{Spin-wave interference patterns recorded by trMOKE at multiple excitation frequencies. In \textbf{a}-\textbf{c}) the FIB irradiated region (grating) is indicated by a semi-transparent red overlay (radius R=30\,$\mu$m, pitch d=12\,$\mu$m), the gray stripe represents the ground line of the CPW. The green circles are the theoretical Rowland circles with a radius R/2, while the radius of red arcs are fitted to the data. Black lines indicate expected diffraction angles, and white arrows point to focal points in the measured data. \textbf{d}-\textbf{f}) Spin-wave intensities extracted along the red arcs in \textbf{a}-\textbf{c}).}
        \label{fig:trmoke}
\end{figure}

We fabricated the designed device and recorded spin-wave interference patterns with trMOKE. The trMOKE images at three different excitation frequencies are shown in Fig. \ref{fig:trmoke}. Diffraction peaks are clearly observed close to the expected diffraction angles, as indicated by black lines. In case of the smallest wavelength the second-order peaks can also be observed ($\frac{2\lambda}{d} < 1$). We found that the peaks are not perfectly focused along the Rowland circle, but an arc can be fitted to the peaks that works well for all three wavelengths. We attribute this to the fact that our grating is much wider than conventional concave gratings ($180^\circ$ instead of a few degrees), and thus the approximations used in the derivation of the Rowland circle do not hold perfectly. This is confirmed by micromagnetic simulations, which also show that focal points are located on an arc with a slightly smaller curvature (see Fig.~\ref{fig:nonlinfmr}\textbf{c}). Another possible cause of deviations is the slight anisotropy of spin waves introduced by a tilt in the external bias field, which can not be fully eliminated in our current experimental setup.

%Measurements in Fig. \ref{fig:trmoke} were performed with single-frequency excitations. Our current measurement setup is not suitable for applying RF signals with multiple frequency components. These measurements did not reveal the nonlinear behavior of the excitation mechanism, apart from the high amplitude oscillations that can be observed behind the grating.

\section*{Generation Mechanism of Spin Waves by Nonlinear Resonance}
\label{sec:SWgeneration}

In most spin-wave devices, the magnetic field of the waveguide is directly responsible for launching the spin waves, as described in \cite{ref:scirep}. The relatively delocalized magnetic field of the waveguide and the localized demagnetizing field of the film edge jointly create a high, periodically changing torque and launch the spin waves \cite{ref:kruglyak}.

We found that in our device an indirect, nonlinear mechanism is dominant for spin-wave excitation. In the design of \cite{ref:scirep} all material in the YIG film is assumed to be removed behind the grating. However, our FIB method is performed as the final fabrication step, after the CPW is already in place. We did not irradiate the total area between the grating and the CPW, only a narrow region, which is wide enough to block spin waves at the designed wavelength to significantly couple through via dipole fields. However, at sufficiently large excitation amplitudes nonlinear effects cause the spin-wave wavelength to increase. This is due to the lowered OOP demagnetization-field component \cite{ref:nonlinear}. The wavelength at sufficiently large amplitudes becomes much larger than the distance between the CPW and the grating. The spin waves that are generated under the CPW reflect back from the back side of the grating, creating a standing-wave pattern. Since the wavelength is much larger than the size of the region between the CPW and the grating, this resembles a homogeneous resonance in that region. The trMOKE images of Fig. \ref{fig:trmoke}\textbf{a}-\textbf{c} already show the high-amplitude region on the left of the grating: it is observable that in this region the colormap is saturated and without apparent pattern, indicating large-amplitude, uniform precession.  

The dipole field of the quasi-homogeneous resonant excitation reaches significantly farther than that of the short-wavelength spin waves, thus it can excite coherent spin waves on the ridges of the grating. This field is in fact much stronger than the magnetic field of the CPW, becoming the dominant effect for spin-wave generation on the grating edge. 

\begin{figure}[ht]    
\centering
    \includegraphics[width=0.99\textwidth]{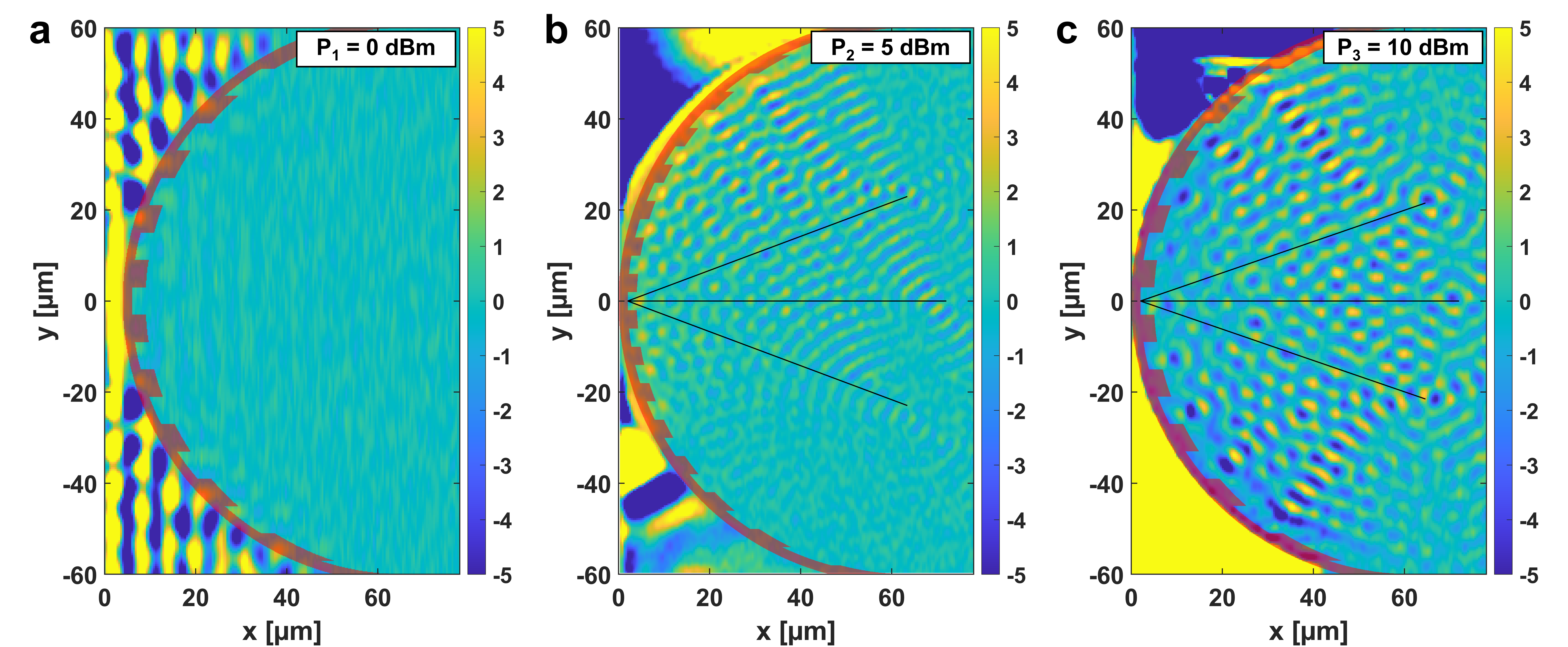}
    \caption{TrMOKE images of miscellaneous gratings expressing the difference in performance with respect to small \textbf{a)}, higher \textbf{b)} and high \textbf{c)} microwave current applied at the input waveguide. The right edge of the CPW ground line is 3 $\mu$m away from grating posterior in each case. The semi-transparent red overlay indicates the FIB-irradiated region, black lines show expected diffraction angles. The grating operation visibly enhances with the level of uniformity in the nonlinear spin wave excitation behind the FIB area.}
    \label{fig:ampdep}
\end{figure}

Figure \ref{fig:ampdep} shows the spin-wave-generation process in more detail. At sufficiently small excitation power ($P_\mathrm{rf} = 0$\,dBm), linear spin waves are excited under the CPW. These small-amplitude, short-wavelength spin waves cannot significantly couple through the FIB-irradiated region. However, the Oersted field of the CPW is not sufficient to create spin waves at the grating edge, that could be detected by our trMOKE apparatus. In Fig. \ref{fig:ampdep}\textbf{b} ($P_\mathrm{rf} = 5$\,dBm) nonlinear behavior is observable behind the grating, but the nonlinear wavelength is not yet long enough to create uniform precession. One can observe a partial interference pattern on the right possibly due to larger uniform standing waves on the top. At $P_\mathrm{rf} = 10$\,dBm excitation power (Fig. \ref{fig:ampdep}\textbf{c}) the resonance almost uniform (apart from the top region being out-of phase), and the interference pattern is complete. 

\begin{figure}[ht]    
\centering
    \includegraphics[width=0.99\textwidth]{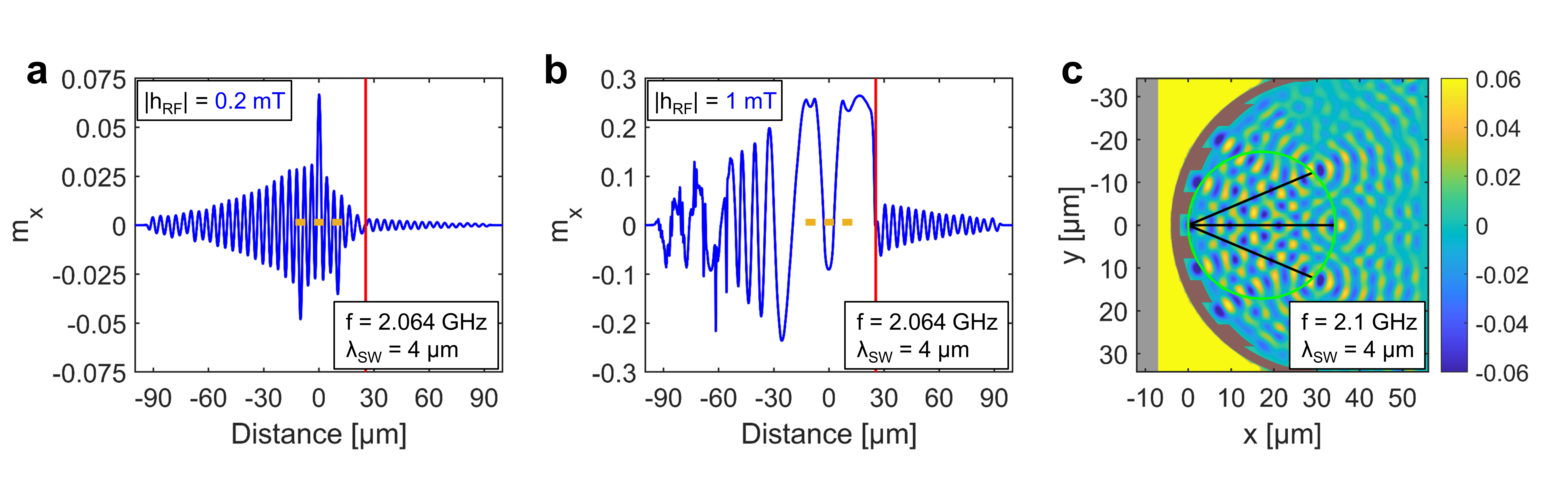}
    \caption{Micromagnetic simulations of nonlinear indirect excitation of spin waves. Yellow rectangles indicate the CPW position and width. The red stripe represents the FIB irradiated region, modelled by zero $\mathrm{M_s}$. \textbf{a)} is a 1D example of linear excitation, while \textbf{b} is a strongly nonlinear case. \textbf{c)} represents a 2D simulation of the experimental scenario in Fig. \ref{fig:trmoke}\textbf{b}.}
    \label{fig:nonlinfmr}
\end{figure}

To gain more insight into the excitation process, we performed both one-dimensional and two-dimensional micromagnetic simulations using mumax3 \cite{ref:mumax}.  These simulations confirmed the behavior that we observed in the experiments (Fig.~\ref{fig:nonlinfmr}). At small excitation fields, spin waves beyond the grating are very small in amplitude (Fig.~\ref{fig:nonlinfmr}\textbf{a}). At a higher (nonlinear) excitation field, however, a uniform standing wave is observed between the grating and the CPW, and coupling through the FIB region is strong (Fig. \ref{fig:nonlinfmr}\textbf{b}). Left from the CPW the wavelength change of the spin waves can be observed as they decay due to magnetic damping. Beyond about 50\,µm propagation self-modulational instability is also causing spikes in the waveform \cite{ref:mod_instability}. 2D simulations also confirmed the operation of the device (Fig.~\ref{fig:nonlinfmr}\textbf{c}). Here the diffraction angles match perfectly the theory, but the position of the focal points are also somewhat behind the Rowland circle. This is the same effect we observe in the experiments. Gratings with larger radius and shorter width would probably not suffer from this deviation, especially at small diffraction angles. However, the position of the peaks is predictable, so this does not affect the usability of the device. 

\def\h{4.7cm} % figure height

An additional, very unusual aspect of this nonlinear excitation can be observed in Fig. \ref{fig:trmoke}\textbf{c}. In case of the direct excitation mechanism one would expect that segments of the grating that are closest to the CPW will excite the highest amplitude waves, since the field of the CPW decays with the lateral distance. However, here we observe exactly the opposite: the strongest "beams" seem to form on the farthest parts of the grating, and amplitudes are less strong in the middle part, which is very close to the CPW (Fig. \ref{fig:ampdep}\textbf{c}). This is because the largest area where homogeneous oscillations can occur are on the sides, where there is enough distance between the CPW and the grating. The larger the area, the higher are the dipole fields, and the higher the excitation on the opposite side of the grating. Thus, the discovered indirect excitation mechanism is very efficient at exciting short-wavelength spin waves on a finely patterned edge at a distance from a CPW. This can be advantageous in applications where a complex wavefront has to be launched. 

A significant limitation of the nonlinear excitation method is that it only works for single-frequency excitations. If multiple frequency components are excited, the uniform standing waves cannot form, moreover nonlinear mixing creates unwanted spectral pollution. If multiple frequencies are present in the excitation signal (as it is often desirable in a spectrum analyzer), the nonlinear method cannot be used, but, as we demonstrated, the proposed method is very effective at creating devices with complex interference patterns at a single frequency.

\section*{Methods}
\label{sec:methods}

\subsection{Sample Fabrication and Characterization}
YIG thin films were deposited on a GGG substrate using rf-magnetron sputtering (we used 100 nm thick films in the grating experiments). Their magnetic properties were evaluated by means of ferromagnetic resonance (FMR), revealing a saturation magnetization of $\mathrm{M_{s}}$=120\,kA/m and a damping constant of $\mathrm{\alpha_{YIG}}=4.4 \times 10^{-4}$. For the excitation of spin waves shorted aluminum CPW antennas were fabricated on top of the YIG film. The antennas were wire bonded to a PCB based CPW with connections to an RF signal generator (Stanford Research SG 386). 

The gratings were fabricated in YIG next to the CPW via FIB irradiation. We used 50\,keV Ga$^+$ ions with a relatively low dose ($10^{15}$ ions/cm$^2$), which is high enough to almost completely destroy the magnetic properties of YIG, but no material is removed. Much higher doses (in case of ion milling) would likely deteriorate the YIG film around the patterned region due to ion scattering, which makes this method more suitable. Sub-micron resolution patterning can easily be achieved (possibly down to 100\,nm in our facility). A similar method was recently described in \cite{ref:fib}, where comparably lower doses were used to change magnetic properties of YIG on film level. Here, we deliberately used higher doses to drastically reduce the saturation magnetization of YIG locally, to create a region which inhibits spin-wave transmission. We found that this method is in effect very similar to actually removing material, as it was proposed in \cite{ref:scirep}. 

\begin{figure}[ht]
    \centering
    \includegraphics[width=0.99\textwidth]{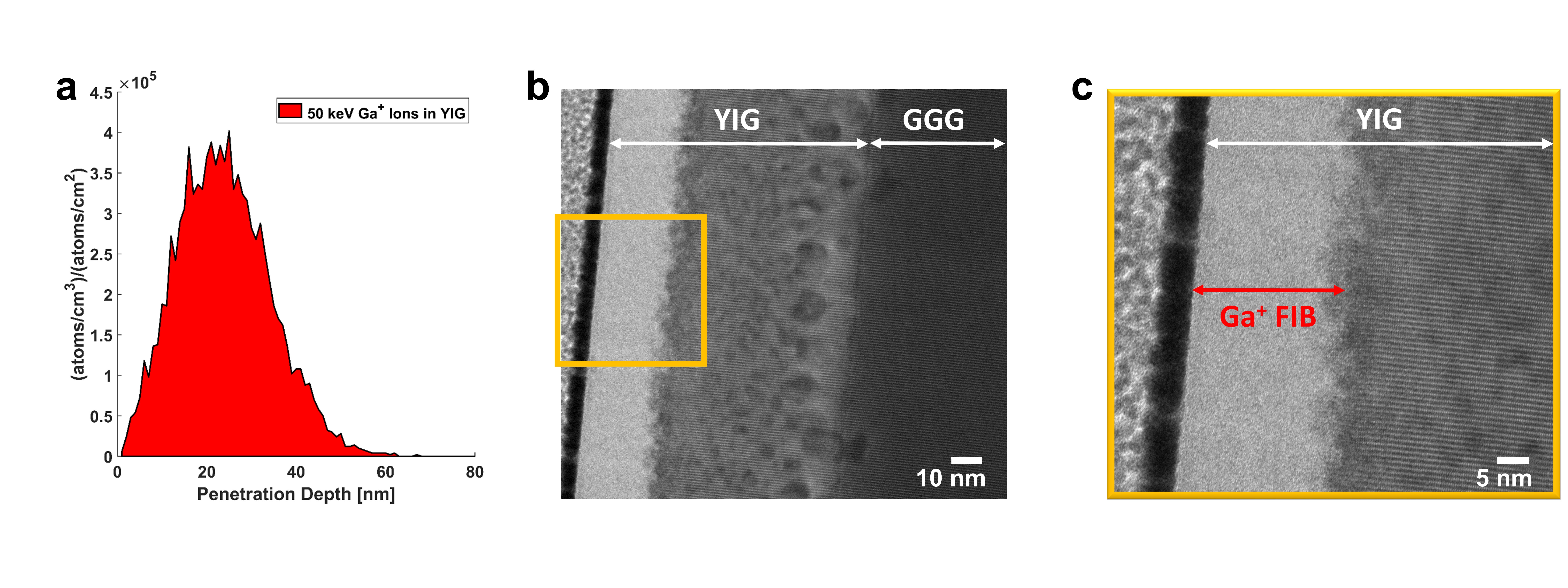}
    \caption{Crystal investigation of the FIB impact in YIG by means of TEM. \textbf{a)} Shows the simulated ion implantation depth for 50~keV Ga$^+$ ions in $Y_{3}Fe_{5}O_{12}$. A cross-sectional image of a 80 nm thick irradiated YIG film is depicted in \textbf{b)}. \textbf{c)} The magnification of the orange square in \textbf{b)} exposes an amorphous toplayer of the thickness expected from the SRIM simulations in \textbf{a)}.}
    \label{TEM}
\end{figure}

The effect of FIB irradiation was also investigated using transmission electron microscopy (TEM). In Fig. \ref{TEM}\textbf{b},\textbf{c} TEM images indicate that the crystalline structure of YIG is completely destroyed down to a depth of approximately 25\,nm, and further significant damage is observable at even higher depths. These results are in good agreement with the previously performed SRIM simulations of our system in Fig. \ref{TEM}\textbf{a}, suggesting a peak implantation depth of 24\,nm.

Ga$^+$ ions are relatively large compared to other frequently used ions such as He$^+$, which explains their low penetration depth and the resulting bilayer formation in YIG. We only had access to Ga$^+$ FIB, but SRIM simulations suggest that higher implantation depths could be achieved with He$^+$ ions, and, with higher doses compared to Ga$^+$, similar modification of YIG properties could be achieved with better uniformity across the film thickness.  

\subsection{Imaging of 2D Spin-Wave Patterns}
To image spin-wave interference patterns we built a time-resolved magneto-optical Kerr microscope (trMOKE). Since the Rowland spectrometer requires isotropic spin-wave propagation, we had to use out-of-plane bias. Thus our trMOKE measures the longitudinal Kerr effect, i.e. it is sensitive to changes in the in-plane magnetization component that lies in the incidence plane of the laser \cite{ref:trMOKE}. We used a ps-laser with 50\,ps pulse width and 405\,nm wavelength (PicoQuant Taiko PDL~M1 with LDH-IB-405 laser head). With this, we can measure spin waves up to approximately 5\,GHz frequency (with 10\,MHz steps) and down to 2\,µm wavelength. We scan through the sample with an XYZ stage with 0.4\,µm resolution. Larger area scans (such as the ones presented in this paper) take about a few hours of measurement time. We use a stroboscopic technique in which the excitation signal is phase locked to the lock-in amplifier and the ps-laser, thus we can extract phase information as well. The amplitude scale is not calibrated, but we estimate that the setup is sensitive to a few percent change in in-plane magnetization.

Currently our setup uses a permanent magnet under the sample for biasing. This makes calibration challenging due to the inhomogeneous field profile. The bias field values in the measurements are approximate values with a few mT uncertainty, and perfect out-of-plane biasing is difficult to achieve. 

\subsection{Micromagnetic simulations}
Micromagnetic simulations were performed in mumax3\cite{ref:mumax}. We used experimental values for parameters where they were available ($\mathrm{M_{s}}$=120\,kA/m and $\mathrm{\alpha_{YIG}}=4.4 \times 10^{-4}$, 100\,nm thickness), and values from literature where we could not directly measure parameters ($\mathrm{A_{ex}}=3.65\times10^{-12}$\,J/m). For discretization we used 30\,nm$\times$30\,nm$\times$100\,nm cells, i.e. approximating the film by a single layer. The lateral cell size is somewhat larger than the exchange length $l_{ex} = \sqrt{2A_{ex}/(\mu_0M_s^2)}\approx{}20$\,nm, but it is still at least a hundred times smaller than the wavelength, and no discrepancies could be observed compared to smaller cell sizes, while the simulation can be completed in a reasonable time. To avoid reflections, in the left and right hand side of the simulation $\mathrm{3\,\mu{}m}$ wide artificial absorbing layers were created using a quadratically increasing damping constant. On the lateral boundaries periodic boundary conditions (single repetition) were used to simulate a long waveguide and avoid energy loss on the sides. The external field was chosen to be 221\,mT, using an analytical dispersion formula to match the measurement wavelength at the given frequency (the external field at the exact position of the sample cannot be measured with sufficient precision in our setup). The FIB irradiation was modeled as a region with ($\mathrm{M_{s}}$=0\,A/m. The field of the CPW was calculated by HFSS, assuming 1 mA current (peak) in the waveguide. The simulation was run for 180\,ns, which was long enough to form a steady interference pattern.

\section*{Conclusions}

Optically inspired magnonic devices represent a promising route to wave-based computing, which are themselves sought after for post von-Neumann computing. To our knowledge, the Rowland spectrometer we demonstrate here is the most complex spin-wave interference device on the micrometer scale. The spin-wave patterns we observe behave remarkably similar to expectations and to the behavior of ideal isotropic waves.

We used FIB irradiation to draw patterns in YIG with nanoscale precision but without removing material. This minimizes the damage to the adjacent YIG areas. The irradiated patterns can influence spin-wave propagation and may also act as wave sources nearby a waveguide. The manipulation of magnetic properties via FIB in other material systems is well-established, but we are not aware of fabricated spin-wave elements in YIG using a similar approach.

Besides demonstrating complex spin-wave patterns in YIG films, we also described a newfound way of creating spin waves via nonlinear resonance, a method that exploits high-amplitude standing-wave oscillations to indirectly excite short-wavelength spin waves with complex wavefronts.

\bibliography{bib}

\begin{thebibliography}{99}

\bibitem{ref:persp} Csaba, G., Papp, Á. and Porod, W., 2017. Perspectives of using spin waves for computing and signal processing. Physics Letters A, 381(17), pp.1471-1476.

\bibitem{ref:mingzhong}	Chang, Houchen, Peng Li, Wei Zhang, Tao Liu, Axel Hoffmann, Longjiang Deng, and Mingzhong Wu. "Nanometer-thick yttrium iron garnet films with extremely low damping." IEEE Magnetics Letters 5 (2014): 1-4.

\bibitem{ref:grundler} Maendl, Stefan, Ioannis Stasinopoulos, and Dirk Grundler. "Spin waves with large decay length and few 100 nm wavelengths in thin yttrium iron garnet grown at the wafer scale." Applied Physics Letters 111, no. 1 (2017): 012403.


\bibitem{ref:snell}	Stigloher, Johannes, Martin Decker, Helmut S. Körner, Kenji Tanabe, Takahiro Moriyama, Takuya Taniguchi, Hiroshi Hata et al. "Snell’s law for spin waves." Physical review letters 117, no. 3 (2016): 037204.

\bibitem{ref:freymann}	Vogel, Marc, Burkard Hillebrands, and Georg von Freymann. "Spin-Wave Optical Elements: Towards Spin-Wave Fourier Optics." arXiv preprint arXiv:1906.02301 (2019).

\bibitem{ref:albisetti}	Albisetti, Edoardo, Silvia Tacchi, Raffaele Silvani, Giuseppe Scaramuzzi, Simone Finizio, Sebastian Wintz, Christian Rinaldi et al. "Optically Inspired Nanomagnonics with Nonreciprocal Spin Waves in Synthetic Antiferromagnets." Advanced Materials (2020): 1906439.

\bibitem{ref:scirep} Papp, Á., Porod, W., Csurgay, Á.I. and Csaba, G., 2017. Nanoscale spectrum analyzer based on spin-wave interference. Scientific reports, 7(1), p.9245.

\bibitem{ref:giovanni} Papp, A., Csaba, G., Dey, H., Madami, M., Porod, W. and Carlotti, G., 2018. Waveguides as sources of short-wavelength spin waves for low-energy ICT applications. The European Physical Journal B, 91(6), p.107.

\bibitem{ref:james} James, J. Spectrograph design fundamentals (Cambridge University Press, 2007

\bibitem{ref:kruglyak} Davies, C. S., and V. V. Kruglyak. "Generation of propagating spin waves from edges of magnetic nanostructures pumped by uniform microwave magnetic field." IEEE Transactions on Magnetics 52, no. 7 (2016): 1-4.

\bibitem{ref:fib} Ruane, W.T., White, S.P., Brangham, J.T., Meng, K.Y., Pelekhov, D.V., Yang, F.Y. and Hammel, P.C., 2018. Controlling and patterning the effective magnetization in Y3Fe5O12 thin films using ion irradiation. AIP Advances, 8(5), p.056007.

\bibitem{ref:mumax} Vansteenkiste, Arne, Jonathan Leliaert, Mykola Dvornik, Mathias Helsen, Felipe Garcia-Sanchez, and Bartel Van Waeyenberge. "The design and verification of MuMax3." AIP advances 4, no. 10 (2014): 107133 see also M.J. Donahue and D.G. Porter Interagency Report NISTIR 6376, National Institute of Standards and Technology, Gaithersburg, MD (Sept 1999)

\bibitem{ref:mod_instability} Wu, Mingzhong, Boris A. Kalinikos, and Carl E. Patton. "Generation of dark and bright spin wave envelope soliton trains through self-modulational instability in magnetic films." Physical review letters 93.15 (2004): 157207.

\bibitem{ref:nonlinear} Wu, Mingzhong. "Nonlinear spin waves in magnetic film feedback rings." Solid State Physics 62 (2010): 163-224.

\bibitem{ref:trMOKE} Arregi, Jon Ander, Patricia Riego, and Andreas Berger. "What is the longitudinal magneto-optical Kerr effect?." Journal of Physics D: Applied Physics 50.3 (2016): 03LT01.

\end{thebibliography}

\section*{Acknowledgements}
The authors are grateful for fruitful discussions and encouragement from Gary Bernstein, Hadrian Aquino (University of Notre Dame), Andrii Chumak (University of Vienna) and Philipp Pirro (TU Kaiserslautern). Many thanks to all staff members at TUM, especially Anika Kwiatkowski. Furthermore, we would like to acknowledge the support of the Central Electronics and Information Technology Laboratory – ZEIT\textsuperscript{lab}.
This work was partially funded by the Deutsche Forschungsgemeinschaft (DFG, German Research Foundation) – Projektnummer (429656450), and by the US National Science Foundation (NSF) with a grant from the SpecEES (Spectrum Efficiency, Energy Efficiency, and Security) program. Adam Papp received funding from the TUM TUFF postdoctoral grant, and from the postdoctoral grant of the Hungarian Academy of Sciences/Eötvös Loránd Research Network (PPD 2019). We are grateful for support from PicoQuant.

%
%\section*{Author contributions statement}
% AP and MK designed the experiments, built the trMOKE, fabricated samples and performed measurements. AP performed micromagnetic simulations. MK, SM, VA and L. Sahin developed YIG fabrication process and performed YIG characterization. SM assisted in the automation of the FIB irradiation process, L. Seitner designed the RF PCB. GC, AP and MK wrote the manuscript. All authors discussed the results and reviewed the manuscript. 

\section*{Additional information}
\textbf{Competing financial interests:} 
The authors declare no competing financial interests.

\end{document}